# Neutrino – Past, Present and Future

## George Marx Memorial Lecture

Univ. Budapest, May 19, 2005

Herbert Pietschmann

Inst. Theor. Physics, Univ. Vienna

## Introduction

Let me begin with my sincere thanks for the opportunity to deliver this Memorial Lecture for George Marx, for we were very close friends during many decades of our lives. When I shall talk about Neutrino Physics, I will select those topics from the vast field in which we both shared deepest interest and about which we had personal discussions.

It was George who some time ago drew my attention to a nice cartoon in an English text book of Electrodynamics:

> *And God said:*
>
> $$\nabla \cdot E = \rho$$
>
> $$\nabla \cdot B = 0$$
>
> $$\nabla \times E = -\frac{\partial B}{\partial t}$$
>
> $$\nabla \times B = j + \frac{\partial E}{\partial t}$$
>
> *- and there was light!*

However, this is a very reductionist view, for light stems from the sun, at least between heaven and earth; but with Maxwell's equations alone the sun cannot shine! Therefore we have to add the following:

> And Wolfgang Pauli said:
>
> *Let there be Neutrinos!*
>
> And Enrico Fermi said:
>
> *Let them interact weakly!*
>
> *And the sun began to shine!*

In fact, when Wolfgang Pauli predicted the Neutrino in 1930, he did not dare to publish it for he feared it might never be detected experimentally. He proposed it in a letter to a conference on radio-activity in Tübingen. And he said to his friend Walter Baade: "Today I have done something which no theoretical physicist should ever do in his life: I have predicted something which shall never be detected experimentally!"[1]

Walter Baade – an astronomer – apparently had great respect for experimentalists and so he bet Pauli that it will one day be detected. And when Reines and Cowan announced the discovery of the neutrino in 1956, Pauli did pay his bet (a case of Champagne)! I wanted to know whether this story is true and at the Neutrino meeting in Aachen I asked Fred Reines (a very close friend of both George and mine) about it. He got very furious and said that yes, it is true, but the Champagne was drank by the theoreticians alone and he and Cowan did not get any drop of it.

But let me not jump forward too fast!

Before the great success of Reines and Cowan it was not very clear, what the best source for neutrinos could be. In a paper[2] by F.G. Houtermans and W. Thirring the authors suggest the sun as source. They estimate the flux of neutrinos from the sun to be $6 \times 10^{10}$ Neutrinos/cm²sec. But in a note added in proof they say: "For technical reasons the publication of this paper has been delayed. Meanwhile it seems that evidence for absorption processes of neutrinos by inverse ß-decay has been obtained by F.Reines and C.L.Cowan."[3]

They refer to the first paper of Reines and Cowan which was criticised for small statistics. The definite acceptance came only after the paper of 1956.[4]

---

[1] For details see G.Marx, Nucl.Phys.B (Proc.Suppl.) **38** (1995) 518
[2] F.G.Houtermans und W.Thirring, Helvetica Physica Acta **27** (1954) 81
[3] F.Reines and C.L.Cowan, Phys.Rev. **92** (1953) 830, 1088.
[4] C.L.Cowan et.al., Science **124** (1956), see also F.Reines and C.L.Cowan, Phys.Rev.**113** (1959)273.

Before we enter into more details, let me give a historical overview of the most important events in Neutrino physics:

<div style="border:1px solid;">

1930:
Wolfgang Pauli: *Prediction of the Neutrino*

1938:
Hans Bethe: *Energy Process of Stars*

1956:
Fred Reines & Clyde Cowan: *Discovery of the Neutrino*

1962:
Lederman, Schwartz, Steinberger et.al.: $\nu_e \neq \nu_\mu$

1964:
John Bahcall: *Calculation of Solar Neutrino Flux*

1967:
Ray Davis: *First Solar Neutrino Experiment* (Cl $\rightarrow$ Ar)

1967:
Bruno Pontecorvo & V. Gribov: *Neutrino Oscillations*

1975:
Martin Perl: *Discovery of 3$^{rd}$ Generation* ($\tau$, $\nu_\tau$)

1987:
*First Observation of a Supernova by Neutrinos*

1998:
Super-Kamiokande: *First Indication of Neutrino Oscillations*

2002:
SNO & KamLAND: *Definite Confirmation of Neutrino Oscillations*

</div>

It is also interesting to see who got the Nobel prize for Neutrino physics:

1938: E. Fermi … NOT for Weak Interactions
1945: W. Pauli … NOT for the Neutrino-Hypothesis
1988: L. Lederman, M. Schwartz, J. Steinberger „for the neutrino beam method and the demonstration of the doublet structure of the leptons through the discovery of the muon neutrino."

1995: F. Reines "for the detection of the neutrino."
2002: R. Davis Jr. and M. Koshiba "for pioneering contributions to astrophysics, in particular for the detection of cosmic neutrinos."

Let us now go through the historical list and see why each of these steps was so fundamental at the time.

Two Kinds of Neutrinos.

Around 1960 the community of particle physicists was faced with a profound problem. The weak decay of the muon was well understood,

$$\mu^- \to e^- + \nu_{(\mu)} + \overline{\nu}_{(e)} \qquad (1)$$

where the indices of the neutrinos are put in brackets because at that time it was not yet known that they are different.
But the seemingly obvious electromagnetic decay was absent!

$$B.R.(\mu \to e + \gamma) \leq 10^{-11} \qquad (2)$$

(The value in Eq.2 is todays best value![5]) Any charged particle-antiparticle pair can be transformed into a photon. Since the neutrino has neither charge nor magnetic moment, it is not possible in this case. However, if the weak interactions are mediated by a charged intermediate boson $W$, there can be inner bremsstrahlung and the decay (2) must be possible, unless it is forbidden by another selection rule, i.e. separate conservation of muon and electron lepton number. In that case, $\nu_e$ and $\nu_\mu$ are different. Thus there was an alternative: Either there exists no intermediate boson $W$, or $\nu_e$ and $\nu_\mu$ are different!
It was obviously a crucial question to know whether all interactions are of Yukawa type or there is an exception, weak interactions of Four-Fermi type! Thus one had to know whether there are 2 types of neutrinos or not.
It was mainly Gilberto Bernardini who pushed with all his personal strength for a neutrino experiment at CERN. To get a feeling for its difficulty, let us recall some basics about neutrino reactions. The total cross-section of a neutrino with energy $E_\nu$ scattering from a target T is

$$\sigma_{tot}(\nu + T \to X) = Const.M_T.E_\nu \qquad (3)$$

where

---
[5] Particle Data Group, Phys.Letters **B592** (2004) 33.

$$\text{Const.} \cong 10^{-38} \text{ cm}^2/\text{GeV}^2. \qquad (4)$$

So for a neutrino in the GeV energy range, scattering from a nucleon, the cross-section is about $10^{-38}$cm² whereas it is $10^{-41}$cm² if the neutrino is in the MeV energy range. (This was the cross section Reines and Cowan had to face in their experiment.)

It would have been a great chance for the young CERN to settle the question of the neutrinos, but CERN aimed at a precision experiment with roughly 1000 events. However, the most crucial point could be answered with just a handful of events; neutrino beams from pions are predominantly $\nu_\mu$ if there are 2 types of neutrinos. Thus their "inverse ß-decay" should produce exclusively muons. Indeed, the experiment was done at the then newborn Brookhaven accelerator and proved the existence of 2 types of neutrinos.[6]

Solar Neutrinos.

It is now more than 40 years since John Bahcall did the first detailed calculation of the flux of neutrinos from the sun. The basic process of Hydrogen burning in the sun is

$$4p \rightarrow \alpha + 2e^+ + 2\nu_e + 26.731 \text{ MeV}. \qquad (5)$$

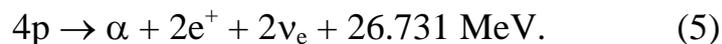

But the detailed reactions are far more complicated as they involve also – among others – $He^3$, $Be^7$, $Li^7$ and $B^8$. The most energetic neutrinos stem from $B^8$ (average energy 7.4 MeV). There is a sharp line from $Be^7$ at $E_\nu=0.862$ MeV. Both these neutrinos could be detected by the inverse ß-reaction

$$\nu_e + Cl^{37} \rightarrow Ar^{37} + e^- \qquad (6)$$

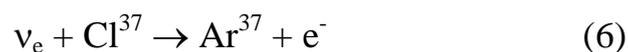

It was Ray Davis who dared to do this experiment. He put a huge tank with 100.000 gallons (~ 400.000 liters) cleaning fluid ($C_2Cl_4$) deep underground in the Homestake mine and washed single atoms of Argon out into a counter.[7] Of course, the expected rate was exceedingly small! The unit used in solar neutrino experiments is called "Solar Neutrino Unit" or SNU. It is defined by

$$1 \text{ SNU} = 10^{-36} \text{ captures/atom.sec} \qquad (7)$$

By 1984, Davis had accumulated enough events to give a solar neutrino flux of 2.1±0.3 SNU.[8] But the prediction of Bahcall was much higher[9], 6-8 SNU depending on the parameters of the solar model. Obviously, this caused a great stir in the community! I remember well the discussion following Davis' talk at one of the Balaton meetings organised by George Marx. The consensus then was that one of the following assertions had to be true:

1. The experiment is wrong
2. The solar model is wrong
3. Nuclear physics is wrong
4. Particle physics is wrong

---

[6] G.Danby et.al., Phys.Rev.Letters **9**(1962)36.
[7] R.Davis Jr. et.al., Phys.Rev.Letters **20**(1968)1205.
[8] R.Davis Jr. et.al., AIP Proc. 123, Steamboat Springs Conf. (1984)1037.
[9] J.N.Bahcall et.al., Rev.Mod.Phys. **54**(1982)767.

At that time, nobody expected the last statement to be the correct one! The situation was so critical that Hans Bethe in his talk at George Marx's Neutrino Symposium 1988 in Boston remarked that he becomes uncertain whether he obtained the Nobel prize for the right reason!
Thus one had to reproduce the experiment with a better threshold. The reaction

$$\nu_e + Ga^{71} \rightarrow Ge^{71} + e^- \qquad (8)$$

has a threshold low enough to observe neutrinos from the primary reaction (5). Two Gallium experiments were set up, "Gallex" at Gran Sasso and "SAGE" (<u>S</u>oviet-<u>A</u>merican-<u>G</u>allium-<u>E</u>xperiment[10]) at the Caucasus. But they also showed a deficit of solar neutrino flux as compared to the solar model prediction which became more and more accurate as time went on.
Finally, the large Cerenkov-Detector "Super-Kamiokande" in Japan came into operation. By means of the elastic reaction

$$\nu_e + e^- \rightarrow \nu_e + e^- \qquad (9)$$

it was not only able to observe solar neutrinos, it could also determine the direction of incidence of the neutrinos because the reaction (9) is strongly forward peaked. Hans Bethe was surely relieved by the result, indeed there were neutrinos from the sun, albeit too few! (I remember vividly how excited George Marx was when he showed me the first graph with neutrino reactions definitely pointing in the direction of the sun!)
At the time of the Neutrino Symposium 2002 in Munich (the last one which George attended!) the ratio of observed neutrino flux to the solar model prediction was:

| | |
|---|---|
| Chlorine Exp. | $0.30 \pm 0.03$ |
| Gallium Exps. | $0.53 \pm 0.03$ |
| Super-Kamiokande | $0.403 \pm 0.013$ |

---

[10] After the decay of the Soviet Union, the experiment was NOT renamed into „RAGE" for "<u>R</u>ussian"!

Before we go to the solution of the solar neutrino puzzle we have to go back to Bruno Pontecorvo's idea of neutrino mixing.

### Neutrino-Oscillations

As early as 1957, Bruno Pontecorvo (a very close friend to George Marx) suggested that neutrinos could oscillate[11] (in analogy to the neutral Kaon system). Mixing of neutrino flavours was suggested by Maki et al.[12] This presupposes that neutrinos of different species have different mass, thus not all of them can be massless. Since the "Standard Model" assumed massless neutrinos[13], this was a courageous step into new physics!

If two species of neutrinos are mixed, we have "mass eigenstates" ($\nu_1$, $\nu_2$, say) differing from "weak eigenstates" ($\nu_e$, $\nu_\mu$). The transition probability of $\nu_l \to \nu_{l'}$ ($l,l'=e,\mu$) is given by

$$P_{ll'} = \sin^2 2\alpha \cdot \sin^2(L/2\lambda), \quad l \neq l' \qquad (10)$$

where $\alpha$ is the mixing angle. For 3 flavours

$$\lambda_{k'k} = \frac{2E}{\Delta m_{k'k}^2} \qquad (11)$$

($k,k'=1,2,3$) with

$$\Delta m_{k'k}^2 = \left| m_{\nu_k}^2 - m_{\nu_{k'}}^2 \right| \qquad (12)$$

and, numerically

$$\frac{L}{\lambda_{k'k}} = 2.54 \frac{\Delta m_{k'k}^2}{eV^2} \frac{L/E}{km/GeV} \qquad (13)$$

Experimentally, one can either look for appearance of a neutrino species at a distance $L$ which was not originally present in the beam of energy $E$ ("appearance experiment"); alternatively, one can measure the thinning of a beam due to oscillation of some of the neutrinos into the other species ("disappearance experiment").

---

[11] B.Pontecorvo, Sov.Phys.JETP **33**(1957)549, **34**(1958)247, **53**(1967)1117.
   V.Gribov and B.Pontecorvo, Phys.Lett. B**28**(1969)493.
[12] Z.Maki, M.Nakagawa, S.Sakata, Progr.Theor.Phys. **28**(1962)870.
[13] D.Haidt and H.Pietschmann, Electroweak Interactions. Landoldt-Börnstein New Series Group I, Vol.10. Springer Verlag Berlin (1988)14.

As stated above, neutrino oscillation requires a mass difference, hence at least one non-vanishing neutrino mass. Direct mass limits are difficult to obtain with great precision. The best limit stems from tritium decay

$$^3H \to {}^3He + e^- + \bar{\nu}_e. \qquad (14)$$

One measures the endpoint of the spectrum in a so-called Kurie plot

$$K(E_e) = \frac{GU_{ud}\sqrt{\xi}}{\pi\sqrt{2\pi}} \sqrt{(E_0 - E_e)^2 - m_\nu^2} \qquad (15)$$

where $\xi$ is the nuclear matrix element and $E_0$ the maximal electron energy $E_e$. Clearly, the intensity of the spectrum is there minimal so one is in need of very many events.
In the course of many experiments some strange phenomena showed up. (George was always most interested in the unexpected!) Until very recently, <u>all</u> experiments gave a negative $m_\nu^2$ while statistically this should only happen in about half the cases! There was also some strange half-year variation in the experiment, called the "Troitsk-effect" which was occasionally interpreted as a huge disk of relic neutrinos through which the earth was travelling twice a year on its orbit. Fortunately all these phenomena have disappeared and the best result on the electron-neutrino mass stems from the two experiments in Mainz and Troitsk; both give a limit of 2.2 eV.
In view of the importance of the neutrino mass, a new experiment – KATRIN – is planned in Karlsruhe. It aims at a sub-eV sensitivity and should reach 0.2 eV by 2013.

### The Third Generation

Much to the surprise of the particle physicists community, a third generation of leptons was discovered by Martin Perl[14] and his collaborators in 1975. Obviously, it was the charged lepton τ which was first seen through the reaction

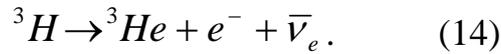

$$e^+e^- \to \mu^\pm + e^{-/+} + \text{missing energy} \qquad (16)$$

The question arose, whether the corresponding neutrino was also of the new 3$^{rd}$ generation or rather one of the known two species. However, these possibilities could soon be ruled out by experiment. It took quite some time, until the third generation neutrino was actually "seen" through its reaction producing a τ-lepton. (In the DONUT-experiment).

---

[14] M.L.Perl et.al.: Phys. Rev. Letters **35**(1975)1489.

(In October 1977, a Triangle-Symposium on <u>Hadron</u> Spectroscopy was organized in Strbske Pleso. George and I were amused by the fact that I gave a talk on "The fifth <u>Lepton</u>" and he on "The sixth <u>Lepton</u>".)

### The Supernova 1987

In February 1987 a Supernova exploded nearby in the Large Magellanic Cloud. For the first time in history, also the neutrino burst from such an event was recorded.[15] Although one could obtain an upper limit[16] on the mass of $\nu_e$, much more information could have been extracted, had the clocks of the two main detectors in Japan and USA been properly synchronized. Unfortunately that was not the case. (I remember my workshop on weak interactions and neutrinos in Santa Fé 1987, where the representative of Kamiokande regretted this with the words "it is very embarrassing since my country is known to export good watches.") They had let a graduate student set the clock of the detector according to his own wrist watch since nobody expected the precise time to be of any relevance!

The big underground detectors were originally developed to discover the decay of the proton. Thus "Kamiokande" was short for "Kamioka <u>N</u>ucleon <u>D</u>ecay <u>E</u>xperiment". When the proton life-time was pushed beyond the capabilities of the detectors and after the historic observation of the Supernova, Kamiokande was renamed to mean "Kamioka <u>N</u>eutrino <u>D</u>etection <u>E</u>quipment".

### Atmospheric Neutrinos[17]

When a cosmic ray particle hits the atmosphere, mainly pions are produced. They decay via $\pi \to \mu + \nu_\mu$ and the muon decays via Eq.(1). From this very simple argument it is clear, that in the cosmic rays twice as many $\nu_\mu$ than $\nu_e$ should be present.[18] Of course there are corrections and there is background (e.g. from produced kaons etc.). However, these corrections can be applied and still the ratio was persistently too low.

A beautiful experiment at Super-Kamiokande measured the ratio of upcoming versus downgoing muons stemming from neutrino interactions. Whilst the downgoing neutrinos had only about 10 km to travel, upcoming had to traverse the whole earth and thus had about 10.000 km to arrive at the detector. Thus they had enough time to oscillate according to eq.(10). Detailed analysis showed that $\nu_\mu$'s oscillated mainly into $\nu_\tau$'s with maximal mixing angle $\sin^2 2\theta = 1$ and a mass difference of $\Delta m^2 = 2.5 \times 10^{-3}$ eV² (see eq.13).

---

[15] K. Hirata et.al.: Phys.Rev.Letters **58**(1987)1490; R.M.Bionta et.al.: Phys.Rev.Letters **58**(1987)1494.
[16] D. Schramm: Proc.Int.Symp.Lepton Photon Hamburg (1987)471.
[17] Atmospheric neutrinos were first observed in India and South Africa:
   C.V. Achar et al.: Phys.Lett. **18**(1965)196; F. Reines et.al.: Phys.Rev.Letters **15**(1965)429
[18] Because of time-dilatation this argument is only correct at lower ν-energies.

Once neutrino oscillations were experimentally suggested, the idea to solve the solar neutrino puzzle by disappearance oscillations of $\nu_e$ became a sound possibility. However, it had to be tested in a positive way also. As long as one could only observe charged current events, disappearance was the only choice. Thus one had to set up a new type of detector which could also observe neutral current neutrino reactions.

### The Sudbury Neutrino Observatory (SNO)

In order to also detect neutral current neutrino reactions, a huge Cerenkov detector filled with 1.000 tons of pure heavy water was built in the Sudbury mine in northern Canada. Deuterium allows for the following neutrino reactions:

$$\nu_l + d \rightarrow \nu_l + n + p \quad l=e,\mu,\tau \qquad (17a)$$

$$\nu_e + d \rightarrow e^- + p + p \qquad (17b)$$

Thus one could observe via eq.(17a) also the total, undiminished neutrino flux from the sun.
It was a great relief and a beautiful success when the results showed indeed that the total neutrino flux from the sun (Neutrino Symp. Munich 2002)

$$\phi_{SNO} = 5.09 +0.44+0.46/-0.43-0.43 \qquad (18a)$$

was in perfect agreement with the theoretical expectation from the Standard Solar Model (SSM)

$$\phi_{SSM} = 5.05 +1.01/-0.81 \qquad (18b)$$

But in order to extract oscillation data similar to the atmospheric case, this was not sufficient. There was still an ambiguity. It had to be resolved by yet another beautiful experiment: KamLAND. It is a 1.000 ton liquid scintillator neutrino detector in Kamioka, Japan, collecting neutrino events from all the surrounding nuclear power plants!
Together, the two experiments gave the best fit for oscillations of solar neutrinos:

$$\Delta m^2 = 8.3 \times 10^{-5} \text{ eV}^2$$
$$\sin^2 2\theta = 0.83 \qquad (19)$$

With three generations of neutrinos and two kinds of oscillations (i.e. two mass differences) this would give a nice picture even if we only know mass differences and not their absolute values. But there is – and we look into the future now – a puzzle still to be resolved.

Sterile Neutrinos?

An oscillation experiment has been performed in Los Alamos, the LSND experiment. When a $\pi^+$ is produced, it decays via $\pi^+ \to \mu^+ + \nu_\mu$ and the positive muon decays via $\mu^+ \to e^+ + \nu_e + \bar{\nu}_\mu$. Thus in a purely positive pion beam there cannot appear any $\bar{\nu}_e$. If they do, they have to stem from appearance oscillations, provided background is carefully subtracted.
The LSND experiment observed just these $\bar{\nu}_e$'s! It is a 4σ result! The trouble is that their measured mass difference is far too large to fit into a three neutrino scheme. But we know that there cannot be a straightforward fourth generation. Already in 1976 I have pointed out[19] that we can extract the number of different neutrino species $N_G$ from the width of the Z-boson by means of

$$\Gamma_Z = \Gamma(Z \to \text{visible}) + N_G \times \Gamma(Z \to \nu\nu). \qquad (20)$$

The value given by the Particle Data Group in 2004 (ref.5) is

$$N_G = 2.994 \pm 0.012 \qquad (21)$$

Thus a fourth neutrino cannot be of the same nature as the other three for it does not couple to the Z-Boson. It is generally called "sterile neutrino" and to me a very ugly creature! Since the LSND result has not yet been independently reproduced, it remains open whether it is correct. In the last years, it was standard practice that the summary of the International Neutrino Conference – Georges child! – was opened with the dry remark "in my summary, I ignore the LSND result!". Still, the question has to be objectively answered! Thus an experiment – MiniBooNE – is under way to check on the LSND result. First results were promised for 2005 but there are some delays because it has to be done very carefully since it is a very important question.
The neutrino physicist community is eagerly waiting for the clarification of this question!
The International Neutrino Conference in Munich (2002) was the last of this series which George was able to attend, to open and to conclude. Two years later, in Paris, I had the honour to remember him[20] at the opening. The series shall go on with undiminished spirit, but we shall miss its father – George Marx – very sincerely.

---

[19] R.Bertlmann and H.Pietschmann: Phys.Rev.**D15**(1977)683.
[20] H.Pietschmann: In Memoriam George Marx. Nucl.Phys.B (Proc.Suppl.)**143**(2005)X.

I am grateful to Walter Grimus for reading the manuscript.